\documentclass[runningheads]{llncs}
\usepackage[T1]{fontenc}
\usepackage{graphicx}
\usepackage[table,dvipsnames]{xcolor}

\usepackage{caption}
\usepackage{cite}
\usepackage{amsmath,amssymb,amsfonts}
\usepackage{hyperref}
\usepackage{pgfplots}
\usepackage{pgfplotstable}
\pgfplotsset{compat=1.18}
\usepackage{pifont}
\usepackage{listings}
\usepackage{csquotes}
\usepackage{algorithmic}
\usepackage{textcomp}

\def\BibTeX{{\rm B\kern-.05em{\sc i\kern-.025em b}\kern-.08em
    T\kern-.1667em\lower.7ex\hbox{E}\kern-.125emX}}

\usepackage{listings}
\usepackage[most]{tcolorbox} 

\usepackage{tabularray}
\newcommand{\mycross}{\ding{55}} 
\newcommand{\mycheck}{\ding{51}}

\newcommand{\mysubsubsection}[1]{\vspace{0.5em}\noindent\textbf{#1}}

\begin{document}
\title{From Chaos to Automation:\\Enabling the Use of Unstructured Data for \\Robotic Process Automation}
\titlerunning{Enabling the Use of Unstructured Data for RPA}
%
\author{Kelly Kurowski\inst{1} \and
Xixi Lu\inst{1} \and
Hajo A. Reijers\inst{1}}
\authorrunning{K. Kurowski et al.}
%
\institute{Utrecht University, Utrecht, the Netherlands\\
\email{\{k.kurowski,x.lu,h.a.reijers\}@uu.nl}}

\maketitle              
\begin{abstract}
The growing volume of unstructured data within organizations poses significant challenges for data analysis and process automation. Unstructured data, which lacks a predefined format, encompasses various forms such as emails, reports, and scans. It is estimated to constitute approximately 80\% of enterprise data. Despite the valuable insights it can offer, extracting meaningful information from unstructured data is more complex compared to structured data. 
Robotic Process Automation (RPA) has gained popularity for automating repetitive tasks, improving efficiency, and reducing errors. However, RPA is traditionally reliant on structured data, limiting its application to processes involving unstructured documents. This study addresses this limitation by developing the UNstructured Document REtrieval SyStem (UNDRESS), a system that uses fuzzy regular expressions, techniques for natural language processing, and large language models to enable RPA platforms to effectively retrieve information from unstructured documents.
The research involved the design and development of a prototype system, and its subsequent evaluation based on text extraction and information retrieval performance. 
The results demonstrate the effectiveness of UNDRESS in enhancing RPA capabilities for unstructured data, providing a significant advancement in the field. The findings suggest that this system could facilitate broader RPA adoption across processes traditionally hindered by unstructured data, thereby improving overall business process efficiency.

\end{abstract}
\section{Introduction}

\begin{figure*}[tb]
    \centering
    \includegraphics[width=\textwidth]{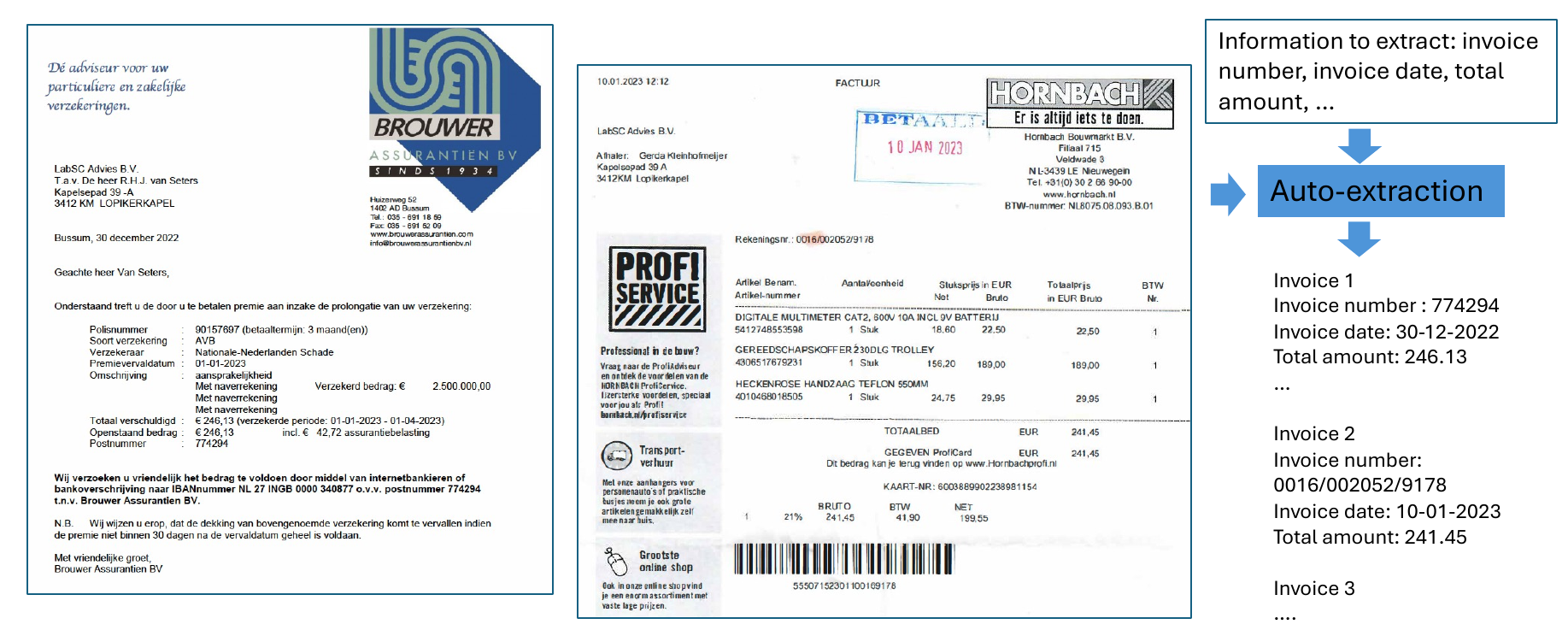}
    \caption{Business-relevant documents, such as invoices, are often captured in unstructured or semi-structured formats (e.g., PDFs or images). Extracting key information like invoice number, total amount, and other essential data is crucial for completing business processes. This figure illustrates this problem using two examples of invoices from \href{https://www.labsc.nl/}{LabSC} with the relevant information extracted.}
    \label{fig:problemExample}
\end{figure*}

Robotic Process Automation (RPA) automates repetitive, rule-based tasks by mimicking user interactions at the user interface (UI) level. This allows organizations to quickly improve efficiency and accuracy without altering underlying systems~\cite{Doguc2022robot}.

However, RPA is most effective when processes involve structured, predictable data~\cite{Costa2022RoboticPA}. In practice, many business documents—such as invoices—are unstructured or semi-structured (e.g., PDFs), making automated processing difficult~\cite{Wewerka2020AUserAcc}. According to IBM, unstructured data now accounts for approximately 80\% of all enterprise data~\cite{ibm_blog}. Fig.~\ref{fig:problemExample} shows an example where business-related documents, such as invoices, are often captured in unstructured or semi-structured formats (e.g., PDFs). Traditional recommendations to address this challenge emphasize the need for companies to ensure that their documents are well organized, structured, and digitally stored~\cite{Costa2022RoboticPA, Enriquez2020Robotic, Syed2020Robotic}. However, it can be argued that RPA platforms lack the necessary capabilities to effectively process unstructured documents, which limits their potential applications.

While Large Language Models (LLMs) offer strong capabilities for extracting information from unstructured texts, they are expensive to train, require substantial data, are prone to hallucinations, and pose sustainability concerns~\cite{Vartziotis2024, Kamath2024}. Sole reliance on LLMs is therefore not viable for many RPA use cases.

We present UNDRESS (UNstructured Document REtrieval SyStem), a framework composed of interchangeable components designed to enable RPA platforms to extract information from unstructured documents. While the individual techniques employed—fuzzy regular expressions, named entity recognition (NER), and LLMs—are well known, their integration in UNDRESS creates a novel and practical solution. Additionally, users can customize UNDRESS by adding their own components or replacing existing ones (e.g., using a different LLM model). Rather than optimizing individual components, our focus is on the overall performance of a pipeline composed of multiple interacting components. While particularly useful for many RPA use cases, UNDRESS is also applicable to a broader range of document processing and information extraction tasks.

We evaluate UNDRESS using two datasets and conduct interviews with RPA developers to validate its practical relevance. Our contribution is a hybrid, extensible system that enables RPA platforms to handle unstructured data more reliably and efficiently. More broadly, we argue that such a hybrid approach supports the development of more robust and sustainable RPA solutions, offers a promising direction for advancing Information Systems Engineering, and encourages the research community to incorporate simpler, more sustainable techniques when building practical systems.

The remainder of this paper is structured as follows: Section~\ref{sec:relatedwork} presents the related work, providing an overview of existing information extraction techniques and approaches. Section~\ref{sec:systemdesign} discusses the system design, including all its modules and components. Section~\ref{sec:evaluation} covers the evaluation of the developed system. In Section~\ref{sec:discussion} we discuss interesting insights. Finally, Section~\ref{sec:conclusion} summarizes our key findings and makes recommendations for future work.

\section{Related Work}





\noindent \textbf{RPA} Most existing RPA methods either assume that there are already structured tables available for RPA solutions to use or assume that there are semi-structured data, such as detailed logs regarding users' mouse clicks and keyboard inputs or form level input~\cite{DBLP:conf/otm/GaoZLA19,DBLP:conf/bpm/MartinezRojasRER24}. These methods then rely mostly on \emph{rule-based} approaches to build bots and can not directly use unstructured data as input for building RPA bots. 

Other works have been studying information retrieval in combination with RPA. They have applied RPA on datasets from specific domains like healthcare \cite{sreekrishna2023systematic}, finance \cite{baidya2021blysis, ling2020intelligent}, human resources \cite{roopesh2021robotic}, etc. 
These works use RPA as a helper tool for retrieving information. For example, they use RPA to monitor the emails and file systems of a particular company, make pre-defined \emph{rules-based} decisions to determine whether the expected information has been found and selected in the extracted text, and download and analyze documents for scanned images. Such solution are data/case specific, as the rules of RPA are hardcoded. Additionally, they do not enhance the RPA technology itself but only use RPA to retrieve information. 

\noindent \textbf{Unstructured Data and Information Retrieval} Data within organizations exists in a wide range of digital formats, including CSV files, Excel spreadsheets, PDFs, emails, images, invoices, financial reports, and audio recordings, among others. This data generally falls into three primary categories: structured, semi-structured, and unstructured data \cite{major2023improving}. 
Unstructured data includes items like emails, images, audio recordings, or PDFs that lack a consistent layout or identifiable elements. This type of data does not use key-value pairs, may include handwritten text, follows a free-flowing structure, and lacks uniformity \cite{major2023improving}.

The data extraction process varies for each category. Although numerous methods are available to extract valuable and useful information from data, no standard approach performs equally well across all scenarios \cite{pustulka2021text}. 

Information extraction techniques are also highly dependent upon the type of document \cite{zaman2020information}. For instance, rule-based extraction methods are ineffective for unstructured documents, as the position of the data is neither predictable nor fixed, but they are very useful for structured data extraction tasks.

While AI-based approaches show potential in autonomously extracting valuable information from unstructured documents, they face challenges in handling diverse layouts. We have found that many studies rely solely on Optical Character Recognition (OCR) for text extraction, which is considered a template-based method due to its limitations in handling inconsistent document layouts \cite{baviskar2021efficient, major2023improving}. It is also recommended to use preprocessing techniques before applying OCR to documents, especially when they are blurry and the text is not very readable. Some case studies even created predefined templates to extract information from documents \cite{gruzauskas2020robotic, ling2020intelligent} which makes the solution highly domain dependent.


To address this issue, there is a recommendation for adopting a template-free, AI-based model. Various document features, such as semantic relationships and positional connections between named entities, can be used to create such a model \cite{baviskar2021efficient}.

\label{sec:relatedwork}


\section{System Design}
\label{sec:systemdesign}



The system has two main modules: the \textit{text extraction} and \textit{information retrieval} modules, each with specific components.

The \textit{text extraction} module includes the Preprocessing component, OCR Engine, Spell Checker, and OpenAI component. It takes an unstructured document as input and outputs extracted text for the next module.

The \textit{information retrieval} module contains Fuzzy Regular Expression, Named Entity Recognition, and OpenAI components. It receives the extracted text and a user query to retrieve the requested information.

\begin{figure}[!t] 
    \centering
    \includegraphics[width=1.0\textwidth]{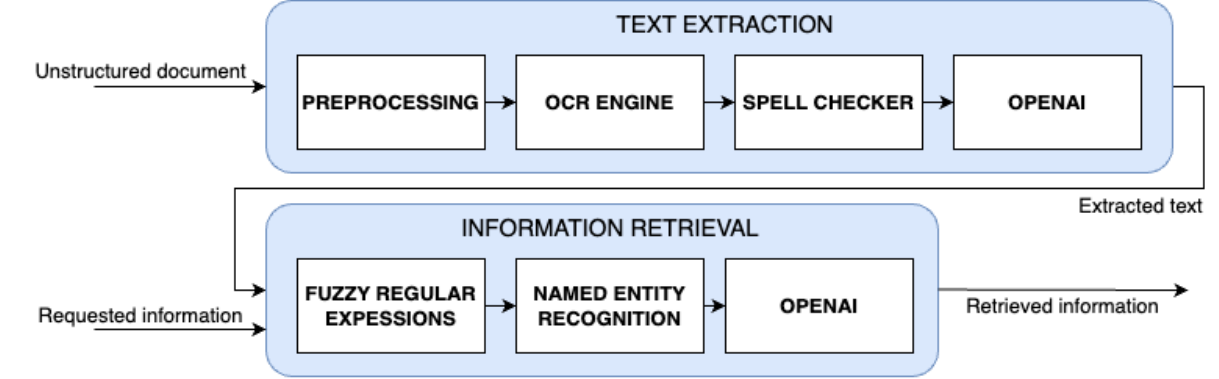} 
    \caption{System Design}
    \label{fig:sysdesign}
\end{figure}

Figure~\ref{fig:sysdesign} shows the system architecture. The following sections describe each component in execution order.

\mysubsubsection{Preprocessing}
To prepare the document for the OCR engine, we initially resize the file to enhance its resolution. Next, we apply a binary threshold to create a clear black and white image, ensuring optimal contrast. Following this, we employ morphological erosion to fill any gaps between letters, enhancing text continuity. Subsequently, we dilate the letters to prevent any distortion or blurring, thus ensuring legibility. Using \href{https://pypi.org/project/opencv-python/}{cv2}, the Python interface for OpenCV (the Open Source Computer Vision Library), we identify contours and create a mask around the text area to isolate it for further processing. Finally, we apply a filter to sharpen the image, enhancing the clarity of the text. With these preprocessing steps complete, the document is now ready to be passed onto the OCR Engine.

\mysubsubsection{OCR Engine}
The Tesseract-OCR Engine is utilized for optical character recognition through the \href{https://pypi.org/project/pytesseract/}{pytesseract library}, an OCR tool for python. Tesseract boasts an accuracy rate of approximately 94\% for identifying numbers and 98\% for identifying letters. The supported file extensions by the developed software include .pdf, .jpg, .jpeg, .img, .tif, and .png.

The detect function from the Python langdetect library identifies the document's language, either Dutch or English, which helps improve OCR accuracy by specifying the language. This step is also essential for applying the spell checker algorithm to the text.

\mysubsubsection{Spell Checker}
We use the Python \href{https://pypi.org/project/pyspellchecker/}{spell checker library} to correct common OCR errors, such as confusing \enquote{l} with \enquote{i}. Words starting with capital letters, often names, are excluded from correction to avoid misprocessing. The spell checker is too simple to handle these cases properly and returns \enquote{None}, which is undesirable since we want to retain those names.

\mysubsubsection{OpenAI - Text Extraction}
Finally, we use an OpenAI API Key to correct and format the extracted text. The model used is GPT-3.5-turbo-1106, which is a specific variant of the Generative Pre-trained Transformer (GPT) model developed by OpenAI. The prompt is formulated according to the best practices for prompt engineering with the OpenAI API\cite{openai_prompt_engineering}, and is as follows:
\definecolor{lightlightergray}{rgb}{0.9, 0.9, 0.9}

\begin{tcolorbox}[colback=lightlightergray, colframe=gray, boxrule=0.4mm, rounded corners, width=\textwidth, enlarge left by=0mm]
\begin{minipage}{0.95\textwidth}\scriptsize
\texttt{prompt = f"""
Correct spelling mistakes and format the following text below. Text: "\{text\}" """}
\end{minipage}
\end{tcolorbox}



\mysubsubsection{Fuzzy Regular Expressions}
For information retrieval, we first let the program search for any fuzzy regular expressions (fuzzy RegEx) that match the user input of information to be retrieved. In this component, we focus on \enquote{static information}, which is the simplest and easiest form of information to retrieve quickly, such as \enquote{e-mail address}, \enquote{phone number} or \enquote{website}, without the need for any document understanding. We formulated a total of nine expressions.
In the code below, an example of a fuzzy RegEx of an e-mail address in Python is given. 
\begin{tcolorbox}[colback=lightlightergray, colframe=gray, boxrule=0.4mm, rounded corners, width=\textwidth, enlarge left by=0mm]
\begin{minipage}{0.95\textwidth}\scriptsize
\begin{verbatim}
fuzzy_regexes = {
    "e-mail": r"\b[A-Za-z0-9._%+-]+@[A-Za-z0-9.-]+\.[A-Z|a-z]{2,}\b" 
}
\end{verbatim}
\end{minipage}
\end{tcolorbox}

Breaking down the expression, it starts with \texttt{\textbackslash{}b}, which denotes a word boundary, ensuring that the match begins at the start of an email address. Next, [A-Za-z0-9.\_\%+-]+ matches one or more occurrences of alphanumeric characters, dots, underscores, percentage signs, plus signs, or hyphens before the \enquote{@} symbol, which is a common pattern in email addresses. After the \enquote{@} symbol, [A-Za-z0-9.-]+ matches one or more occurrences of alphanumeric characters, dots, or hyphens, which represent the domain name. \. matches a literal dot before the top-level domain, which can consist of two or more characters specified by [A-Z|a-z]\{2,\}. Finally, \texttt{\textbackslash{}b} marks the end of the email address, ensuring that it is matched as a whole word.

This regular expression is fuzzy because it matches a wide range of e-mail formats, including those with varying characters and lengths, allowing it to identify addresses even with slight deviations from the standard pattern.

\mysubsubsection{Named Entity Recognition}
When a fuzzy RegEx match is not identified, the program proceeds with NER. This component mainly focuses on extracting information such as organizations, companies, persons, cities, places, etc. 

We used the \enquote{nl\_core\_news\_lg} model from the spaCy library, which boasts the highest named entity precision rate (0.79) among all the Dutch models offered by spaCy. Additionally, we used the \enquote{en\_core\_web\_trf} model for English named entities, which also achieves the highest named entity precision rate (0.90) among the English models offered by spaCy.

\mysubsubsection{OpenAI - Information Retrieval}
When no information is extracted or found by the NER component, we utilize an OpenAI API Key to retrieve the specific information that the user is seeking from the text. We generate the following prompt, adhering to the guidelines provided by OpenAI: 
\begin{tcolorbox}[colback=lightlightergray, colframe=gray, boxrule=0.4mm, rounded corners, width=\textwidth, enlarge left by=0mm]
\begin{minipage}{0.95\textwidth}\scriptsize
\texttt{prompt = f"""Extract \{user\_input\} from the following text below.\\
Text: "\{text\}" \{user\_input\}: """    }
\end{minipage}
\end{tcolorbox}






This component is the final part of the Information Retrieval module. By this point, all the information that the user wishes to extract should have been retrieved by the software, if it has not already been obtained by preceding components.


\section{Evaluation Experiment}
\label{sec:evaluation}
Our evaluation has three main objectives: first, to assess the text extraction module against the ground truth; second, to measure the information retrieval module’s effectiveness on various fields (e.g., invoice amount, date) using accuracy, recall, and precision; and third, to explore practical applicability through interviews with five RPA developers. We tested the two modules separately on distinct datasets. Below, we describe the datasets, metrics, and results.

\subsection{Datasets}
We used two datasets containing unstructured documents for text extraction and information retrieval.

\mysubsubsection{Real Invoices}
The first dataset consists of 100 purchase invoices, all with very different layouts, from \href{https://www.labsc.nl/}{LabSC}, a Dutch company specialized in comprehensive services for laboratory facilities and equipment. LabSC provides on-site inspections of fume hoods, maintenance for lab furniture, and consulting for laboratory setups and air handling systems.

\mysubsubsection{Generated Resumes}
The second dataset consists of 400 resumes generated by us using a Python script with the \href{https://pypi.org/project/reportlab/#description}{Reportlab} and \href{https://faker.readthedocs.io/en/master/}{Faker} libraries. Faker provides randomized names, addresses, and emails, while labels for these entities vary to reflect unstructured document practices. Document sections and titles are randomized in both position and naming (e.g., a section labeled \enquote{Skills} might appear as \enquote{Competencies}, \enquote{Talents}, \enquote{Skillset}, or \enquote{Strengths}). Layouts and content vary, with the script choosing from 70 occupations, 50 work experiences, and 30 academic backgrounds to closely mimic unstructured document diversity. The script generates different PDF layouts with adjusted margins, randomly chosen title fonts and sizes, varied heading styles, and decorative horizontal lines for section separation. 
The script can be found in the \texttt{generate\_fake\_resumes.py} file in the \texttt{Data} folder.\footnote{\url{https://github.com/Kelly-Kurowski/UnstructuredDataExtractionTool}.}

\subsection{Evaluation Metrics}
To evaluate the \emph{text extraction} module for both datasets, we used the Jaccard Similarity Index, a quantitative metric that measures the ratio of common words to total unique words in the extracted set and ground-truth set. The closer the result is to 1, the more similar the two sets are. Each word has an equal weight in this metric, meaning that it does not consider content or semantic relationships. The metric is particularly suited for evaluating whether UNDRESS extracts all the text from the documents. For each document in the dataset, we compute the Jaccard Similarity Index and report the average over all the documents. 

To determine the accuracy of the \emph{information retrieval} module, first, we counted the total number of correct outputs—those that matched the ground truth from the document. Then, we computed the accuracy by dividing the number of correct outputs by the total number of tested documents, which in this case was always 40.

We also report precision and recall for the \textit{information retrieval} module, aiming to assess 
the system's efficacy in retrieving relevant information.
%
Labels that were correctly retrieved and matched exactly one correct answer were classified as True Positives (TP). If a label is retrieved and returned multiple times, even if the label is included in the ground truth, only one of them was labeled TP, the rest were labeled as False Positives (FP). Additionally, if there was a single incorrect label extracted, it was also labeled as a False Positive (FP). Finally, if no match was found when there should have been information in the document, it was labeled as a False Negative (FN).
We calculate $\text{Precision} = \frac{TP}{TP+FP}$, and  $\text{Recall} = \frac{TP}{TP+FN}$.



\subsection{Results - Text Extraction}
\mysubsubsection{Resumes} 
Because the resume dataset is artificially generated, it consists of text-based PDFs, allowing for the selection and copying of all text within them. We simply use the extract\_text function from the Python pdfminder library to extract the text from each document and use them as our ground truth. Subsequently, we compared the ground truth text with the results obtained from our \emph{text extraction} module using the Jaccard Similarity Index.

We experimented with various prompts in the GPT-3.5 model to determine which one yielded the best results. The first prompt yielded an average Jaccard Similarity Index of 0.90. We observed that 10\% of the extracted text from our solution did not align with the original text due to additional symbols accompanying the words. For instance, the Text Extraction module included \enquote{satisfaction} in its set, while the PDF extractor included \enquote{satisfaction,}. 


For the second prompt, we removed \enquote{and format the following text below} leaving: \enquote{Correct spelling mistakes in the following text if there are any.} The temperature was set to 0.1 for deterministic responses, favoring conservative corrections. This approach yielded an average Jaccard Similarity Index of 0.92.

For the third prompt, we added a negation, despite following the best OpenAI API prompt practices. The prompt reads: \enquote{Correct spelling mistakes in the following text if there are any; do NOT format the text}. The third test did not yield a significant improvement, with an average score of 0.91, which was better than the first test but worse than the second. 

Table 1 shows the OpenAI prompts with their respective scores. Finally, we used the prompt with the best avarage Jaccard Similarity Index in combination with standard NLP methods by applying tokenization and lemmatization. The resulting average Jaccard Similarity Index was 0.99, achieving a significantly higher score.

\mysubsubsection{Invoices}
The invoice dataset contains text in PDFs that is not selectable, preventing it from being copied. Consequently, the Python pdfminer library cannot be used to create the ground-truth sets, as it may not extract all the text. To address this limitation, we utilized Adobe Acrobat Pro to recognize the non-selectable text. We verified manually if the text extracted by Adobe Acrobat Pro did not contain any mistakes.

We found that the extracted text sometimes confused certain letters; for instance, a \enquote{c} was mistaken for a \enquote{(} symbol. These errors were also corrected manually. We did this manual refinement on 100 documents to create the \emph{ground-truth} texts. 

After obtaining the ground-truth texts, we ran the exact same four tests as for the \emph{Resumes} dataset. 
The prompt \enquote{Correct spelling mistakes in the following text if there are any} yielded an average Jaccard Similarity Index of 0.71. We also tested the prompts \enquote{Correct spelling mistakes and format the following text below} and \enquote{Correct spelling mistakes in the following text if there are any, do NOT format the text}, this gave an Index of 0.69 and 0.71 respectively.

In the fourth test, we applied the same rules used for the resume dataset. Combining these rules with the second prompt (which had the highest Jaccard Similarity Index) improved the score to an average of 0.81.

In Table 1, the reader can find a concise overview of the four tests that were run with the specified prompt
per dataset, excluding the mitigation of special symbols. Note that employing prompt engineering techniques could have led to even better results, but this was not the main objective of this research.

\begin{table}[t]
\centering
 \resizebox{.9\textwidth}{!}{
\begin{tabular}{p{10cm} p{1.5cm} p{1.5cm}}
\textbf{OpenAI Prompt} & \multicolumn{2}{p{3cm}}{\raggedright \textbf{Average Jaccard Similarity Index}}\\
\hline
 & Resumes& Invoices\\
\hline
1) Correct spelling mistakes and format the following text below  & 0.90&  0.69\\
\hline
2) Correct spelling mistakes in the following text if there are any  & 0.92&  0.71\\
\hline
3) Correct spelling mistakes in the following text if there are any, do NOT format the text  & 0.91&  0.71\\
\hline
4) Used Prompt (2) and applied the tokenization and lemmatization rules  & 0.99&  0.81\\
\end{tabular}
}
\label{tab:results}
\caption{Results of OpenAI Prompts on Datasets}
\end{table}

\subsection{Results - Information Retrieval}

\mysubsubsection{Resumes}
In the resume dataset, we extracted personal information as done in the works of Roopesh et al. \cite{roopesh2021robotic}, Barducci et al. \cite{barducci2022AnEnd}, Wosiak \cite{wosiak2021automated} and Pudasaini et al. \cite{pudasaini2022application}, such as the candidate's name, address, e-mail, and phone number. Furthermore, we were interested in identifying the candidate's educational background, (current) job position, known languages, and skills. The retrieval techniques chosen for each field are done in a sequential manner (i.e., first try Fuzzy regex, then NER, then LLM), as discussed in Section~\ref{sec:systemdesign}. 

Table 2 lists the results for each field. The field \enquote{Name} achieved an accuracy score of 83\%. The fields \enquote{Address}, \enquote{E-mail}, \enquote{Phone number} and \enquote{Language} each had a perfect score of 100\%. Additionally, \enquote{Job title} demonstrated strong performance with a 95\% accuracy score, and \enquote{Education}, yielded an impressive 98\% score. For skills, we considered \enquote{Soft skills} and \enquote{Hard skills} in the resume. The corresponding precision and recall for this dataset can be found in Table 2. 



\begin{table}[h!]
  \centering
  \begin{minipage}{0.45\textwidth}
        \centering
        \resizebox{1.1\textwidth}{!}{
            \begin{tabular}{ l l c c c }
            
            \textbf{Field} & \textbf{Technique} & \textbf{Accuracy} & \textbf{Precision} & \textbf{Recall} \\
            \hline
            Name & NER & 0.83 & 0.83 & 1.00 \\
            \hline
            Address & Fuzzy regex & 1.00 & 1.00 & 1.00 \\
            \hline
            E-mail & Fuzzy regex & 1.00 & 1.00 & 1.00 \\
            \hline
            Phone number & Fuzzy regex & 1.00 & 1.00 & 1.00 \\
            \hline
            Education & LLM & 0.98 & 0.98 & 1.00 \\
            \hline
            Hard skills &  LLM & 0.93 & 0.93 & 1.00 \\
            \hline
            Soft skills  &  LLM & 0.88 & 0.88 & 1.00 \\
            \hline
            Job title &  LLM & 0.95 & 0.95 & 1.00 \\
            \hline
            Language & NER & 1.00 & 1.00 & 1.00 \\
            
            \end{tabular}
        }
        \label{tab:accuracy_precision_recall_fields}
        \caption{Accuracy, Precision, and Recall for the resume dataset}
  \end{minipage}%
  \hspace{0.05\textwidth} 
  \begin{minipage}{0.45\textwidth}
        \centering
        \centering
        \resizebox{1.1\textwidth}{!}{
            \begin{tabular}{l l  c c c}        
            \textbf{Field} & \textbf{Technique} & \textbf{Accuracy} & \textbf{Precision} & \textbf{Recall} \\
            \hline
            Invoice number & LLM & 0.90 & 0.90 & 1.00 \\
            \hline
            Invoice date & Fuzzy regex & 0.80 & 1.00 & 0.77 \\
            \hline
            Total amount  &Fuzzy regex & 0.65 & 0.72 & 0.87 \\
            \hline
            IBAN & Fuzzy regex & 0.80  & 0.87 & 0.87 \\
            \hline
            Seller & LLM & 0.83 & 0.83 & 1.00 \\
            
            \end{tabular}
        }
        \label{tab:accuracy_precision_recall}
        \caption{Accuracy, Precision, and Recall for the invoice dataset}
  \end{minipage}
\end{table}

\mysubsubsection{Invoices}
For the invoice dataset, we took the intersection of the terms evaluated by Keturis et al. \cite{kerutis2022intelligent}, Arslan \cite{Arslan2022End}, and Rohaime et al.\cite{rohaime2022integrated}, who also used an invoice dataset. This resulted in the terms \enquote{Invoice number}, \enquote{Invoice date}, and \enquote{Total amount}. 
Additionally, we retrieved the international bank account number (\enquote{IBAN}), evaluated by Arslan \cite{Arslan2022End}, and the name of the company that sent the invoice (\enquote{Seller}), which was evaluated in the work of Kerutis et al. \cite{kerutis2022intelligent}.

The extraction accuracy for various fields was measured across 40 documents. The field \enquote{Invoice number} achieved a 0.90 accuracy rate, while \enquote{Invoice date} was correctly extracted 0.80 of the time. The accuracy for \enquote{Total amount} stood at 0.65. The \enquote{IBAN} field was accurately identified 0.80 of the time, and the \enquote{Seller} field had an extraction accuracy of 0.83. The corresponding precision and recall can be found in Table 3. 
A comparison of our system components with other works in the context of RPA can be found in Table 5.




\mysubsubsection{LLM versus Fuzzy Regex and NER}  We also compared the performance of LLM to the performance of Fuzzy Regex or NER, since it is possible to use the LLM component for each field. Figures~\ref{fig:accuracyLLMresume} and~\ref{fig:accuracyLLMinvoice} show the results for the \emph{Resume} and \emph{Invoice}, respectively. 
For most of the fields (e.g., Name, total amount, IBAN), LLM seems to outperform NER or Fuzzy regex. Interestingly, when the name of the field can be interpreted in different ways, LLM (ChatGPT3.5) seems to be confused. In our case, when we ask to retrieve \enquote{Language}, the LLM component retrieves the language of the document, instead of the languages that the candidates speak, achieving a significantly lower accuracy than NER. The user could use a different LLM model to potentially mitigate this specific error.
Another observation worthy of mentioning may be that for the invoice date field, the Fuzzy Regex component achieves a higher precision (i.e., 1.00) than the LLM component (i.e., 0.89). It should be noted that the LLM component achieves a better recall (0.94 vs. 0.8).

\begin{figure}[h!]
    \centering
    \begin{minipage}{0.45\textwidth}
        \centering
        \resizebox{\textwidth}{!}{
        \begin{tikzpicture}
            \begin{axis}[
                ybar,
                bar width=3pt, 
                enlarge x limits=0.05, 
                ylabel={Accuracy},
                symbolic x coords={Name, Address, E-mail, Phone number, Education, Hard skills, Soft skills, Job title, Language},
                xtick=data,
                x tick label style={rotate=45, anchor=east, font=\scriptsize}, 
                legend style={
                    at={(0.5,-0.6)}, 
                    anchor=north, 
                    legend columns=3, 
                    font=\scriptsize
                }, 
                nodes near coords,
                nodes near coords align={vertical},
                ymin=0,
                ymax=1.00,
                width=7.5cm, 
                height=4cm,
                yticklabel style={
                    /pgf/number format/fixed, 
                    /pgf/number format/precision=1,
                    /pgf/number format/zerofill
                } 
            ]
            \addplot[color=Red, fill=Red] coordinates {
                (Name, 1.00) (Address, 1.00) (E-mail, 1.00) 
                (Phone number, 1.00) (Education, 0.98) 
                (Hard skills, 0.93) (Soft skills, 0.88) 
                (Job title, 0.95) (Language, 0.23) 
            };
            \addlegendentry{LLM}

            \addplot[color=blue, fill=blue] coordinates {
                (Address, 1.00) (E-mail, 1.00) 
                (Phone number, 1.00) 
            };
            \addlegendentry{Fuzzy regex}

            \addplot[color=ForestGreen, fill=ForestGreen] coordinates {
                (Name, 0.83) (Language, 1.00)
            };
            \addlegendentry{NER}
            \end{axis}
        \end{tikzpicture}
        }
        \caption{Accuracy results per extraction method for resume fields}
        \label{fig:accuracyLLMresume}
    \end{minipage}%
    \hspace{0.05\textwidth}  
    \begin{minipage}{0.45\textwidth}
        \centering
        \resizebox{\textwidth}{!}{
        \begin{tikzpicture}
            \begin{axis}[
                ybar,
                bar width=3pt,
                enlarge x limits=0.05, 
                ylabel={Accuracy},
                symbolic x coords={Invoice number, Invoice date, Total amount, IBAN, Seller},
                xtick=data,
                x tick label style={rotate=45, anchor=east, font=\scriptsize},
                legend style={
                    at={(0.5,-0.6)},
                    anchor=north, 
                    legend columns=3, 
                    font=\scriptsize
                },
                nodes near coords,
                nodes near coords align={vertical},
                ymin=0,
                ymax=1.00,
                width=7.5cm, 
                height=4cm, 
                yticklabel style={
                    /pgf/number format/fixed, 
                    /pgf/number format/precision=1,
                    /pgf/number format/zerofill
                } 
            ]
            \addplot[color=Red, fill=Red] coordinates {
                (Invoice number, 0.90) (Invoice date, 0.85) 
                (Total amount, 0.88) (IBAN, 0.88) 
                (Seller, 0.83)
            };
            \addlegendentry{LLM}

            \addplot[color=blue, fill=blue] coordinates {
                (Invoice number, 0.03) (Invoice date, 0.80) 
                (Total amount, 0.65) (IBAN, 0.85) 
            };
            \addlegendentry{Fuzzy regex}

            \addplot[color=ForestGreen, fill=ForestGreen] coordinates {
                (Seller, 0.00)
            };
            \addlegendentry{NER}
            \end{axis}
        \end{tikzpicture}
        }
        \caption{Accuracy results per extraction method for invoice fields}
        \label{fig:accuracyLLMinvoice}
    \end{minipage}
\end{figure}

\subsection{Results - Qualitative Evaluation}

The primary goal of the qualitative evaluation was to assess the value of UNDRESS for RPA developers in their work activities. To achieve this, we demonstrated the system to five RPA developers: three from a large academic hospital in Utrecht, the Netherlands, and two from an RPA consultancy in Amsterdam, the Netherlands. A concise overview of the developers and their work experience is provided in Table 4.

\begin{figure}[h]
    \centering
    \begin{minipage}{0.45\textwidth}
        \resizebox{1.1\textwidth}{!}{
            \begin{tabular}{c c p{3em}}
                \textbf{Developer} & \textbf{Organization} & \textbf{Experience (Years)} \\
                \hline
                Developer 1 & UMCU$^{1}$, Utrecht, the Netherlands & 4 \\
                \hline
                Developer 2 & UMCU, Utrecht, the Netherlands & 4 \\
                \hline
                Developer 3 & UMCU, Utrecht, the Netherlands & 1 \\
                \hline
                Developer 4 & Tacstone$^{2}$, Amsterdam, the Netherlands & 5 \\
                \hline
                Developer 5 & Tacstone, Amsterdam, the Netherlands & 2 \\
                \hline
            \end{tabular}
        }
        \captionof{table}{RPA Developers}
        \label{tab:rpa-developers}
        \vspace{0.3em}{\scriptsize \textit{1. UMCU is a large academic hospital. 2. Tacstone is an RPA Consultancy.}}
    \end{minipage}
    \hfill
    \begin{minipage}{0.45\textwidth}
        \resizebox{1.1\textwidth}{!}{
            \begin{tikzpicture}
                \begin{axis}[
                    ybar,
                    bar width=7pt,
                    enlarge x limits=0.25,
                    ylabel={Frequency},
                    symbolic x coords={Ease of use, Applicability, Novelty},
                    xtick=data,
                    x tick label style={align=center, font=\small},
                    legend style={at={(0.5,-0.2)}, anchor=north,legend columns=-1, font=\scriptsize},
                    nodes near coords,
                    nodes near coords align={vertical},
                    ymin=0,
                    ymax=5,
                    width=7.5cm,
                    height=5cm,
                    ]

                    \addplot[color=red, fill=red] coordinates {(Ease of use, 0) (Applicability, 0) (Novelty, 0)};
                    \addlegendentry{Strongly Disagree}

                    \addplot[color=orange, fill=orange] coordinates {(Ease of use, 0) (Applicability, 1) (Novelty, 0)};
                    \addlegendentry{Disagree}

                    \addplot[color=Goldenrod, fill=Goldenrod] coordinates {(Ease of use, 1) (Applicability, 3) (Novelty, 1)};
                    \addlegendentry{Neutral}

                    \addplot[color=ForestGreen, fill=ForestGreen] coordinates {(Ease of use, 4) (Applicability, 1) (Novelty, 4)};
                    \addlegendentry{Agree}

                    \addplot[color=blue, fill=blue] coordinates {(Ease of use, 0) (Applicability, 0) (Novelty, 0)};
                    \addlegendentry{Strongly Agree}

                \end{axis}
            \end{tikzpicture}
        }
        \caption{Survey results for RPA Developers by statement}
        \label{fig:survey}
    \end{minipage}
\end{figure}

After the demo session, there was room for questions and an open discussion. Next, each developer was asked to fill in a 5-point rating scale (i.e., strongly disagree, disagree, neutral, agree, strongly agree) on the following statements, which focused on ease of use, applicability, and novelty:

\begin{enumerate}
\item The graphical user interface of UNDRESS is intuitive and easy to navigate \textit{(Ease of use)}.
\item UNDRESS could be used to automate many more business processes than are currently possible \textit{(Applicability)}.
\item UNDRESS is a significant addition to the current activities on RPA platforms \textit{(Novelty)}.
\end{enumerate}

Finally, we asked all developers why they provided the answers they did and what they would suggest for further improvement.

\autoref{fig:survey} illustrates the results of the survey. Four developers found the system easy to navigate, while one developer had a neutral opinion. This developer suggested adding annotations to clarify the system's function and use of multiple arguments in the extraction field. The others mentioned that the errors and gray text in the field boxes are helpful for intuitive use.


Three out of five developers expressed a neutral opinion on the second statement. The developers highlighted the phrase \enquote{many more processes} from the statement, noting that several industrial solutions already facilitate information extraction from documents. One developer mentioned \href{https://www.klippa.com/en/home-en/}{Klippa}, a company that automates finance and compliance processes, focusing on unstructured documents such as invoices, IDs, and declarations, but not many more types of documents. A general tool for handling diverse unstructured documents across departments is still lacking. Most developers agreed that while information extraction is feasible through various industrial vendors, UNDRESS significantly simplifies the process.

Regarding the third statement, four out of five developers agreed that it is a valuable addition to current RPA platform activities. Two developers praised the sequence in which information is extracted, noting that it progresses from simpler methods (fuzzy regex and NER) to more complex ones (AI). They remarked, \enquote{Most solutions focus solely on AI, which can sometimes be unnecessarily complex when a label could be extracted more easily with regex. I find this solution very effective in that regard.} One developer mentioned frustration with UiPath's Document Understanding library, noting that the AI sometimes complicates the process and makes it difficult to extract a label, when a simpler regular expression could achieve the exact same result more effectively.

\begin{table*}[]
\label{tab:comparison}
\resizebox{\textwidth}{!}{%
\begin{tabular}{ll|llll|llll|}
\cline{3-10}
 &   & \multicolumn{4}{c|}{\textbf{Text extraction}}     & \multicolumn{4}{c|}{\textbf{Information retrieval}}         \\ \hline
\multicolumn{1}{|l|}{\textbf{Work}}                                  & \textbf{Year} & \multicolumn{1}{l|}{\textbf{OCR}} & \multicolumn{1}{l|}{\textbf{LLM-Extraction}} & \multicolumn{1}{l|}{\textbf{Preprocessing}} & \textbf{Spell Checker} & \multicolumn{1}{l|}{\textbf{Fuzzy}} & \multicolumn{1}{l|}{\textbf{NER}} & \multicolumn{1}{l|}{\textbf{LLM-Retrieval}} & \textbf{Data set} \\ \hline
\multicolumn{1}{|l|}{UNDRESS}                                       & 2024          & \multicolumn{1}{l|}{\mycheck}     & \multicolumn{1}{l|}{\mycheck}                & \multicolumn{1}{l|}{\mycheck}               & \mycheck               & \multicolumn{1}{l|}{\mycheck}       & \multicolumn{1}{l|}{\mycheck}     & \multicolumn{1}{l|}{\mycheck}               & Invoice, resume   \\ \hline
\multicolumn{1}{|l|}{Kerutis et al.  ~\cite{kerutis2022intelligent}} & 2022          & \multicolumn{1}{l|}{\mycheck}     & \multicolumn{1}{l|}{\mycross}                & \multicolumn{1}{l|}{\mycross}               & \mycross               & \multicolumn{1}{l|}{\mycross}       & \multicolumn{1}{l|}{\mycross}     & \multicolumn{1}{l|}{\mycross}               & Invoice           \\ \hline
\multicolumn{1}{|l|}{Roopesh et al.  ~\cite{roopesh2021robotic}}     & 2021          & \multicolumn{1}{l|}{\mycheck}     & \multicolumn{1}{l|}{\mycross}                & \multicolumn{1}{l|}{\mycross}               & \mycross               & \multicolumn{1}{l|}{\mycross}               & \multicolumn{1}{l|}{\mycheck}             & \multicolumn{1}{l|}{\mycross}                       & Resume            \\ \hline
\multicolumn{1}{|l|}{Rohaime et al.  ~\cite{rohaime2022integrated}}  & 2022          & \multicolumn{1}{l|}{\mycheck}     & \multicolumn{1}{l|}{\mycross}                & \multicolumn{1}{l|}{\mycheck}               & \mycross               & \multicolumn{1}{l|}{\mycross}       & \multicolumn{1}{l|}{\mycross}     & \multicolumn{1}{l|}{\mycross}               & Invoice           \\ \hline
\end{tabular}%
}
\caption{Text extraction and information retrieval: comparison of different works}
\end{table*}

\section{Discussion}
In this section, we discuss the results and describe our insights into the performance of the system and its modules.

Our text extraction module demonstrated competitive performance compared to state-of-the-art solutions. For the resume dataset, it achieved a Jaccard Similarity Index of 0.99, while for the invoice dataset, the score was 0.81. While Kerutis et al. \cite{kerutis2022intelligent} reported that 83\% of invoices were processed correctly, the lack of clarity regarding their metric makes direct comparisons challenging. Our solution did outperform Kerutis et al. \cite{kerutis2022intelligent} on fields such as \enquote{Invoice number} (0.90 vs. 0.68) and \enquote{Seller field} (0.83 vs. 0.33). Rohaime et al. \cite{rohaime2022integrated} achieved perfect scores for key invoice fields, but their dataset was more structured than ours.

Specific fields in the invoice dataset presented unique challenges. For instance, while our solution achieved perfect precision (1.00) for the \enquote{Invoice date} field, its accuracy (0.80) was lower due to variations in label alignment, such as when labels were top-aligned rather than left-aligned. Similarly, errors in the \enquote{IBAN} field (0.80 accuracy) often resulted from character misinterpretations, such as confusing \enquote{0} with \enquote{O}. Additionally, multiple \enquote{Total amount} labels in some invoices led to the extraction of redundant values, highlighting a limitation of fuzzy regular expressions. These results demonstrate the \textit{trade-offs in using different techniques} and underscore the need for \textit{field-specific adjustments} to improve accuracy.

The performance difference between resume and invoice datasets likely stems from document characteristics. Resumes have fewer numerical values, making them simpler to extract, while invoices are more complex due to their heavy use of numbers, increasing the chance of extraction errors. Some invoices were scanned, introducing noise and blurriness, while the resume dataset was synthetically generated, ensuring cleaner input. Despite these challenges, the results are fair, as real-world resumes are typically well-readable.

Our approach demonstrates the \textit{versatility} of combining LLM, fuzzy regular expressions, and NER. For instance, Figure 3 illustrates that fuzzy regular expressions achieve accuracy comparable to LLMs for fields such as \enquote{Address}, \enquote{Email}, and \enquote{Phone number}, while NER even outperforms the LLM in extracting specific fields, like \enquote{Language}. This highlights that \textit{a solution based solely on LLMs is not necessarily required}.

The use of LLMs and a spell checker algorithm was also a distinctive feature of our approach, enabling us to process datasets in two different languages. In addition, we are the first to use an LLM integrated hybrid approach in  RPA context. The proposed system could benefit RPA by processing hundreds of documents automatically. Moreover, the retrieved information could be directly useful for real-life applications, such as automating financial reporting or streamlining HR processes. 

However, LLMs introduced unique challenges. For instance, the OpenAI component often formatted text despite explicit prompts to avoid doing so, leading to inconsistencies. Interestingly, even when explicitly instructed not to format the text, there was no significant improvement, \textit{indicating the difficulty of controlling the behavior of LLM}. By mitigating special symbols in the text, we achieved a 0.07 improvement in the Jaccard Similarity Index for the resume dataset and a 0.10 improvement for the invoice dataset. These findings also highlight the importance of \textit{balancing LLM capabilities with simpler, more controllable methods to avoid unnecessary complexity}.


\label{sec:discussion}

\section{Conclusion}

We developed and evaluated UNDRESS, a system that enables RPA to work effectively with unstructured documents. The architecture centers on two key components: text extraction and information retrieval. Our primary contribution lies in the novel integration of fuzzy regular expressions, NER, and LLMs. This hybrid approach outperforms methods focused on a single technique, particularly across diverse document types.

Whereas prior work often limits scope to a single domain—such as invoices, where regular expressions may suffice—we show that this does not generalize to other formats like resumes, which require more semantic interpretation, favoring NER. Our results validate the effectiveness of our method on both synthetic resume and invoice datasets, achieving perfect accuracy (1.00) for personal information (e.g., addresses, emails, phone numbers), and high scores for job titles (0.95), education (0.98), invoice numbers (0.90), and sellers (0.83).

Importantly, UNDRESS demonstrates that high accuracy is achievable without full reliance on LLMs, which are expensive to train, data-intensive, and environmentally burdensome. By combining simpler, interpretable methods with optional LLM components in a modular framework, our system offers a customizable, efficient, and sustainable solution for real-world RPA applications. With this work, we aim to demonstrate to the research community that LLMs are not the only viable path to high-performing document automation. Instead, based on the evaluation results, a balanced hybrid strategy can perform just as well.

A qualitative evaluation with RPA developers revealed that the system would significantly help them in their jobs by simplifying the extraction of information from unstructured data. The developers acknowledged that the system effectively simplifies the extraction of information from unstructured data. They appreciated the progression from simpler methods to more complex AI techniques, noting that UNDRESS avoids unnecessary complexity.

For future work, we propose enhancing label-entity alignment through spatial analysis, improving handling of duplicate or ambiguous labels, and refining the system’s decision-making about which techniques to apply. The ability to seamlessly switch between methods—based on context or performance—underscores the strength of our hybrid design and opens up promising opportunities for further optimization.

\vspace{0.5em}
\noindent \textbf{Acknowledgement} We would like to thank Inge van de Weerd for her valuable contribution to the paper.
\label{sec:conclusion}




\bibliographystyle{splncs04}
\bibliography{literature}
\end{document}